\documentclass[prd,twocolumn,showpacs,preprintnumbers]{revtex4}
\pdfoutput=1
\usepackage{amsfonts,amsmath,amssymb}
\usepackage{graphicx}
\usepackage{color}
\usepackage{natbib}
\pdfoutput=1

\begin{document}

\newcommand{\apjl}{Astrophys. J. Lett.}
\newcommand{\apjs}{Astrophys. J. Suppl. Ser.}
\newcommand{\aap}{Astron. \& Astrophys.}
\newcommand{\aj}{Astron. J.}
\newcommand{\araa}{Ann. Rev. Astron. Astrophys. } 
\newcommand{\mnras}{Mon. Not. R. Astron. Soc.}
\newcommand{\jcap}{JCAP}
\newcommand{\pasj}{PASJ}
\newcommand{\pasa}{Pub. Astro. Soc. Aust.}

\title{Constraints on  frequency-dependent violations of  Shapiro delay from  GW150914}
\author{Emre  O. \surname{Kahya}$^1$} \altaffiliation{E-mail: eokahya@itu.edu.tr}
\author{Shantanu  \surname{Desai}$^2$} \altaffiliation{E-mail: shntn05@gmail.com}

\affiliation{$^{1}$Department of Physics, Istanbul Technical University, Maslak 34469 Istanbul, Turkey}
\affiliation{$^{2}$Ronin Institute, Montclair, NJ 07043, USA }

\begin{abstract}

On 14th September 2015, a transient gravitational wave (GW150914)  was  detected by  the two LIGO detectors at Hanford and Livingston from the coalescence of  a  binary black hole system  located at a distance of about 400 Mpc. We point out that GW150914  experienced a   Shapiro delay due to the gravitational potential of the mass distribution along the line of sight of  about 1800 days.   Also, the  near-simultaneous arrival  of gravitons over a frequency range of about 200 Hz within a 0.2 second window  allows us to constrain  any violations of Shapiro delay and Einstein's equivalence principle between the gravitons at different frequencies. From the calculated Shapiro delay and the observed duration of the signal, frequency-dependent  violations of the equivalence principle for gravitons are constrained to  an accuracy of  $\mathcal{O}(10^{-9})$.  
\end{abstract}

\maketitle

\section{Introduction}

In 1964, I. Shapiro~\cite{Shapiro} argued that the  round-trip time of a   radar signal to the inner planets of our solar system experiences a delay  caused by the  non-zero gravitational potential of the Sun (if it is close to  the line of sight), as  a consequence of Einstein's equivalence principle (EEP). This delay  is referred to  in the literature as    ``Shapiro delay'' and has been measured precisely in the solar system for about five decades, allowing very stringent tests of  general relativity (GR)~\cite{Will} as well as in binary pulsars, where it has been used as an astrophysical probe to measure neutron star masses~\cite{Demorest}. Following the detection of neutrinos from SN~1987A~\cite{IMB,Kamioka}, it was pointed out  that  the neutrinos from SN~1987A also experience  a Shapiro delay due to the gravitational potential of the intervening matter along the line of sight~\cite{Longo,Krauss,Franson}. The value for the delay ranged from  one to six  months for different models of the  galactic gravitational potential~\cite{Longo,Krauss}. 
   The near-simultaneous arrival of photons and neutrinos from this core-collapse supernova confirmed that the Shapiro delay for neutrinos is same as that for photons to within 0.2-0.5\%~\cite{Longo,Krauss}.

About eight years ago, it was pointed out that a gravitational wave (GW)  will also undergo  Shapiro delay due to the line of sight   gravitational potential~\cite{Desai}. In other words, gravitational waves also gravitate,  which implies that the speed of gravitational waves in our universe, which is  filled with matter is (very) slightly smaller than the speed of light, contrary to the standard lore  that they travel at the speed of light.  Of course, this delay is negligible compared to  the total travel time assuming that there is no mass between the source and the Earth. In the case of a GW signal with an electromagnetic counterpart, one can use the relative Shapiro delay between GWs and photons/neutrinos to rule out or confirm alternate theories of gravity which dispense with the need for dark matter, also known as ``dark-matter emulators''~\cite{Desai,Kahya08,Kahya10,Desai15}. Most recently it was also pointed out that in case of an observed GW signal,  in addition to ``multi-messenger'' tests of Shapiro  delay, one can also constrain frequency-dependent violations of Shapiro delay using the fact that gravitons of different frequencies arrive nearly simultaneously~\cite{Sivaram,Gao}.

In September 2015, the LIGO detectors started a new science run called O1 with a  sensitivity of about 1500-2000 Mpc to binary  black hole coalescence assuming optimal orientation~\cite{LIGOdetchar}. On 14th September 2015 at 09:50:45 UTC just before the start of O1, a GW signal (designated as GW150914)  with statistical significance of 5.1$\sigma$ and a  combined signal-to-noise ratio of 24 was detected using data from the two LIGO detectors in Hanford and Livingston. The inspiral part of the signal lasted for about 0.2 seconds in the frequency range from 35-250 Hz.  From the observed morphology, the signal is consistent with   a binary black hole (BBH)  merger with the masses of two companions equal to $(36 \pm 5)  M_{\odot}$ and $(29 \pm 4)  M_{\odot}$, and   the estimated luminosity distance  equal to   about 400   Mpc~\cite{LIGO}. From the observed signal, many  tests of GR (including the GW  speed) have already been carried out~\cite{ligogr,sawicki}.  This is a watershed moment in the history of astronomy and opens a brand new observational window into the universe. 

This GW signal was followed up by a large number of  electromagnetic (EM)  and neutrino followup teams.  Although no statistically significant  EM or neutrino signal was seen at the time of writing, this is not surprising if the event is  due a BBH merger.  However, there was
a weak transient source around 50~keV detected  by the Fermi Gamma-Ray Burst Monitor about 0.4 seconds after the GW signal,
with a false alarm probability of 0.0022~\cite{gbm}. If it is definitively established that this Fermi detection  is associated with GW150914, then DM emulator models are effectively ruled out. However,  an acid test of these models should be possible  in case the next detected GW signal comes from neutron star mergers or core-collapse supernovae, which have guaranteed EM/neutrino counterparts.  In this work we use only the observed multi-frequency GW signal to set a bound on any frequency (energy)-dependent violations of Shapiro delay. Since the Shapiro delay also gets modified by a non-zero mass~\cite{Bose} (in addition to the dispersion relation), one could obtain independent bounds on the graviton mass complimentary 
to those obtained in ~\cite{ligogr}.


\section{Estimated Shapiro delay for GW150914}
Wei et al ~\cite{Wei} (and references therein),  have  pointed out that  for any astrophysical messenger (photons, gravitational waves, neutrinos)  seen across a broad frequency spectrum, one can use the relative time difference between the carriers at multiple frequencies to constrain frequency-dependent Shapiro delay violations, which in turn allows us to set a stringent limit on any violations of EEP. This technique has been applied to EM observations from Fast Radio Bursts~\cite{Wei},  gamma-ray bursts~\cite{grb}, and also TeV blazars~\cite{blazar}. Currently the best limits are obtained for photons from Fast Radio Bursts   of $\mathcal{O}(10^{-9})$~\cite{Wei}. They also proposed a similar test for GWs in case of an observed detection~\cite{Gao}. Following their suggestion, we now apply this method to set limits on any frequency-dependent violations of Shapiro delay for GW150914.

GW150914 was detected at a  luminosity distance of $410^{+160}_{-180}$  Mpc. In order to calculate the total Shapiro delay we first consider the delay due to the Milky Way. The dominant effect will come from its dark matter distribution, which was calculated for a Navarro-Frenk-White dark matter profile, and the estimated delay is approximately   300 days at a distance of 400 kpc~\cite{Desai,Kahya10}. After exceeding the virial radius, the delay follows a logarithmic behavior as a function of distance. The Shapiro delay for the Milky Way using a  Schwarzschild  metric and treating the total gravitating mass as a point source can be written as~\cite{Longo,Krauss,Wei}:  

\begin{equation}
\Delta t^{\mathrm  MW}_{\mathrm  shapiro}=(1+\gamma) \frac{GM_{\mathrm  MW}}{c^3} \ln\left(\frac{d}{b}\right)\; ,
\end{equation}
where $\gamma$ is the parameterized post-newtonian (PPN) parameter, $b$ is the impact parameter, and $d$ is the distance to the source. For  $M_{\mathrm MW} =1\times 10^{12}M_\odot$, $d=400$~kpc,  
$b=8$~kpc, and $\gamma=1$ (assuming GR is correct), this equation gives  $\Delta t^{\mathrm MW}_{\mathrm shapiro} \sim 445$ days. Therefore, this delay is about the same as that    estimated previously considering only the dark matter potential~\citep{Desai,Kahya10}. For an order of magnitude estimate of  the total delay from the Milky way, we use the value of  300 days from our previous calculations. Our assumption of treating the galactic potential as a point source improves as the distance increases, since the galaxies behave point-like at large distances. One would   get a logarithmic enhancement at a distance of  400 Mpc compared to 400 kpc, which increases the delay by about a factor of three. Moreover,  we would also need to consider the total number of Milky way like  galaxies that the GWs pass through. One can add the  combined surface area of all the galaxies within a sphere of radius 400 Mpc and divide that by the surface area of a sphere of radius 400 Mpc. Alternately, one can also consider a cylindrical line of sight, whose  surface area  is determined  from the   galaxy virial radius and height by the distance to the source, and  then divide its volume by  the total volume of a sphere having  radius  equal to the distance to the source, and then multiply by the total number of galaxies within this spherical volume. After doing this,  one gets a factor of $ \sim (r_{vir}/400 \mathrm{Mpc})^2\times N_{tot}$, where  $r_{vir}$ is the virial radius equal to  $250$ kpc~\cite{Klypin}, and $N_{tot}$ is the total number of galaxies within 400 Mpc  equal to  $3\times 10^{6}$~\cite{peebles}. This  contributes an additional factor of two. Taking all of these into account, the total calculated Shapiro delay is equal to  $6\times 300$ days  or about 1800 days.

Given the large distance to the source of GW150914, one should also consider  cosmological effects on the Shapiro delay, which would increase our estimated value. Nusser~\cite{Nusser} made a statistical analysis of this effect and found that it dominates the Milky Way-induced delay by several orders of magnitude for a distance of  1500 Mpc. For a distance of 400 Mpc the effects will not be that large but  should be included for a more accurate estimate.

We should also point out that there is a large  uncertainty in the angular position  of GW150914 (90 \% confidence level region covers about 600 deg.$^2$~\cite{skylocation}). However, since we have assumed homogeneity to get an order of magnitude estimate, the direction of the source should not change the delay by an appreciable amount. The angular dependence  of the Shapiro delay was worked for the  Milky Way only and shown to be less than 10\%~\cite{Kahya10}.

Once the Shapiro delay due to the gravitational potential of the matter distribution along the line of sight is calculated, one can 
use the fact that the GW signal over a  bandwidth  of about 200 Hz consists of multi-frequency gravitons, which arrived within 0.2 
seconds to set a conservative limit on the frequency-dependent violations of EEP for gravitons using the procedure outlined in ~\cite{Wei}. Analogous to photons, if we  define a  PPN parameter for GWs as   $\gamma_{gw}$, the  frequency (energy)-dependent violation of EEP is given by: $|\gamma_{gw} (250 Hz)- \gamma_{gw} (35 Hz)| < 2.6 \times 10^{-9}$.

\section{Conclusions}
The LIGO detection of GW150914 has opened a new observational window into the universe and already provided a plethora
of information on binary black hole mergers, strong field gravity, GW speed, astrophysical populations of binary black holes, etc.~\cite{LIGO,sawicki}. Here, we point out that  GW150914 has gravitated due to the potential of the intervening mass distribution
along the sight. We do an order of magnitude estimate of this  Shapiro delay, which  is approximately equal to 1800 days. Following the suggestion of Wu et al~\cite{Gao}, if we treat the observed multi-frequency GW signal as coming from  different gravitons, we can constrain EEP using the fact that the gravitons arrived within a 0.2 second window. The  violation of EEP for gravitons in terms of the  PPN parameter $\gamma_{gw}$ is given by  $|\gamma_{gw} (250 Hz)- \gamma_{gw} (35 Hz)| < 2.6 \times 10^{-9}$.  More tests of relative Shapiro delay between GWs and photons/neutrinos should be possible with the next set of LIGO/VIRGO detections from sources with EM counterparts.


\begin{acknowledgements}
We thank Richard Woodard and Yizhong Fan for useful correspondence. EOK   acknowledges support from Tubitak Grant Number: 112T817. EOK would also like to dedicate this paper to the memory of Steven Detweiler.
\end{acknowledgements}
\bibliography{eep}
\end{document}